# Measuring the Polarization of a Rapidly Precessing Deuteron Beam

Z. Bagdasarian,[1,2] S. Bertelli,[3] D. Chiladze,[1,2] G. Ciullo,[3] J. Dietrich,[2] S. Dymov,[2,4]
D. Eversmann,[5] G. Fanourakis,[6] M. Gaisser,[2] R. Gebel,[2] B. Gou,[2,7] G. Guidoboni,[3] V. Hejny,[2]
A. Kacharava,[2] V. Kamerdzhiev,[2] A. Lehrach,[2] P. Lenisa,[3] B. Lorentz,[2] L. Magallanes,[8,9]
R. Maier,[2] D. Mchedlishvili,[1] W.M. Morse,[10] A. Nass,[2] D. Oellers,[3] A. Pesce,[3] D. Prasuhn,[2]
J. Pretz,[5] F. Rathmann,[2] V. Shmakova,[2,4] Y.K. Semertzidis,[11,12] E.J. Stephenson,[13,*]
H. Stockhorst,[2] H. Ströher,[2] R. Talman,[14] P. Thörngren Engblom,[3,15] Yu. Valdau,[16,17]
C. Weidemann,[3] and P. Wüstner[18]

[1]High Energy Physics Institute, Tbilisi State University, GE-0186 Tbilisi, Georgia
[2]Institut für Kernphysik, Forschungszentrum Jülich, 52425 Jülich, Germany
[3]University of Ferrara and/or INFN, 44100 Ferrara, Italy
[4]Laboratory for Nuclear Problems, Joint Institute for Nuclear Research, RU-141980 Dubna, Russia
[5]III. Physikalisches Institut B, RWTH Aachen University, D-52056 Aachen, Germany
[6]Institute of Nuclear Physics NCSR Demokritos, GR-15310 Aghia Paraskevi, Athens, Greece
[7]Institute of Modern Physics, Chinese Academy of Sciences, Lanzhou 730000, China
[8]Ludwig Maximilians Universität München, Geschwister Scholl Platz 1, 80539 Munich, Germany
[9]Universitätsklinikum Heidelberg, Im Neuenheimer Feld 450, 69120 Heidelberg, Germany
[10]Brookhaven National Laboratory, Upton, New York 1973, USA
[11]Center for Axion and Precision Physics Research, Institute for Basic Science,
291 Daehak-ro, Yuseong-gu, Daejeon 305-701, Republic of Korea
[12]Department of Physics, KAIST, Daejeon 305-701, Republic of Korea
[13]Indiana Unversity Center for Spacetime Symmetries, Bloomington, IN 47405 USA
[14]Cornell University, Ithaca, New York, 14850 USA
[15]Department of Physics, Royal Institute of Technology, SE-10691, Stockholm, Sweden
[16]Petersburg Nuclear Physics Institute, 188300 Gatchina, Russia
[17]Helmholtz-Institut für Strahlen- und Kernphysik, Universität Bonn, Nussallee 14-16, D-53115 Bonn, Germany
[18]Zentralinstitut für Engineering, Elektronik, and Analytik - Systeme der Elektronik (ZEA-2), Forschungszentrum Jülich, 52425 Jülich, Germany

PACS: 29.20.db, 29.27.Bd, 29.27.Hj, 29.85.Ca

**Abstract:**

This paper describes a time-marking system that enables a measurement of the in-plane (horizontal) polarization of a 0.97-GeV/c deuteron beam circulating in the Cooler Synchrotron (COSY) at the Forschungszentrum Jülich. The clock time of each polarimeter event is used to unfold the 120-kHz spin precession and assign events to bins according to the direction of the horizontal polarization. After accumulation for one or more seconds, the down-up scattering asymmetry can be calculated for each direction and matched to a sinusoidal function whose magnitude is proportional to the horizontal polarization. This requires prior knowledge of the spin tune or polarization precession rate. An initial estimate is refined by re-sorting the events as the spin tune is adjusted across a narrow range and searching for the maximum polarization magnitude. The result is biased toward polarization values that are too large, in part because of statistical fluctuations but also because sinusoidal fits to even random data will produce sizeable magnitudes when the phase is left free to vary. An analysis procedure is described that matches the time dependence of the horizontal polarization to templates based on emittance-driven polarization loss while correcting for the positive bias. This information will be used to study ways to extend the horizontal polarization lifetime by correcting spin tune spread using ring sextupole fields and thereby to support the feasibility of searching for an intrinsic electric dipole moment using polarized beams in a storage ring. This paper is a combined effort of the Storage Ring EDM Collaboration and the JEDI Collaboration.

## I. Introduction

It has been proposed [1] that a storage ring may be used to search for an intrinsic electric dipole moment (EDM) on the charged particles in the circulating beam. The measurement would begin with a beam whose polarization is held against rotation in the ring plane by a suitable choice of ring electric and magnetic fields. For the proton EDM search where the anomalous magnetic moment is positive, it appears easiest to build a storage ring with all electric elements where the choice of $p = 0.7$ GeV/c for the beam momentum makes it possible to match the rotation rates of the velocity and the in-plane polarization. For the deuteron EDM search where the anomalous magnetic moment is small and negative, a crossed magnetic and electric (pointed outward) field is required to match the rotation rates. In both cases, there is an electric field in the particle frame that is directed toward the center of the ring. For the deuteron, most of this field arises from the $\gamma \vec{v} \times \vec{B}$ term in the transformation to the particle frame. It is several times larger than the imposed electric field. If an EDM along the particle spin axis is present, the polarization will precess out of its initial longitudinal orientation and acquire a vertical component.

Since any small vertical polarization component may not be completely canceled at the beginning, the experiment requires measuring the *change* in the vertical component between the beginning and the end of the beam storage time. This change in the vertical component is proportional to the EDM multiplied by the time integral of the longitudinal polarization during the measurement. A possible way to sample the beam polarization while minimizing systematic errors has been discussed by Brantjes [2]. Additionally, the longitudinal component of the beam polarization must remain large throughout the measurement time despite the tendency toward spin decoherence by differential precession in the ring plane. For a statistical sensitivity to the EDM of about $10^{-29}$ e·cm within a year of data acquisition, a balance of requirements would be a sensitivity of the polarimeter to a change of $10^{-6}$ in the vertical polarization component combined with a longitudinal polarization lifetime of 1000 s.

A study has been initiated with the Cooler Synchrotron (COSY) at the Forschungszentrum Jülich [3] to investigate the use of various tools, especially sextupole fields, to extend the horizontal polarization lifetime of a polarized deuteron beam. But without adding electric field plates to each bending dipole in the ring, such a study must confront the task of measuring the beam polarization as it rotates at a high rate (~ 120 kHz) in the ring dipole magnetic field. This paper describes a time-marking system that records the clock time of each scattering event that makes up the polarimeter data set and uses this time to unfold the rotating polarization. Thus a polarization-induced down-up count rate asymmetry can be recovered from the data by identifying and separating those events that occur when the polarization is predominantly sideways to the beam direction, producing a sensitivity to this component of the rotating polarization.

The precession rate of the spin in the ring plane relative to the rotation of the beam velocity is given by the spin tune, $\nu_S = G\gamma$, where $G$ is the particle's anomalous magnetic moment (also denoted as $a$ for leptons) and $\gamma$ is the relativistic factor. For a coasting beam, the spread in particle momenta gives rise to a spread in the values of $\gamma$ and a widening distribution of spins in the plane of the ring. For a typical momentum distribution width $(\Delta p/p)$ of $10^{-4}$ and a ring revolution time of one μs, depolarization times can be as short as tens of ms. This effect can be countered to first order by bunching the beam, thus requiring that all particles maintain the same average revolution time. Variations in $\gamma$ then become periodic with the synchrotron frequency and their contribution to depolarization largely averages out over time. This leaves the leading contributors to depolarization being second-order processes such as path lengthening due to betatron oscillations and other terms containing $(\Delta p/p)^2$.

The magnitude of these second-order processes is subject to adjustment through the application of sextupole fields since the quadratic rise of the sextupole field away from its center can be made to match the varying sizes of second-order driving terms (transverse $\Delta x$ and $\Delta y$ deviations and $\Delta p/p$) through the local size of the betatron oscillation (in both transverse directions) and the access to momentum spread in places of large horizontal dispersion. A demonstration of the ability to reduce the spread of spin tunes using combinations of sextupole corrections thus becomes a feasibility requirement for the design of an EDM storage ring.

This paper includes a description of the procedures used to extract the horizontal polarization magnitude as a function of time during the beam storage cycle. From such measurements, the depolarization time may be inferred. This paper discusses the features of the time-marking system and a technique for interpreting the data. The results of systematic studies of the effects of sextupole fields on the spin coherence time will be discussed in subsequent publications.

Section II describes the setup of the experiment and shows some types of measurements made possible with the new time-marking system. Section III provides a description of the main systematic errors. Section IV discusses how the spin coherence time is defined using various horizontal polarization time functions and how the time dependence may be modeled assuming that the spin tune spread is mainly a consequence of betatron oscillations. This section also deals with the correction for the main systematic error, positive bias in the horizontal polarization magnitude. Information is also provided throughout on the typical performance of the time-marking system and the COSY ring. The conclusions appear in Section V.

## II. Experimental Details

*A. Beam and polarimeter setup*

This study was made using a beam of 0.97-GeV/c polarized deuterons circulating in the Cooler Synchrotron (COSY) located at the Forschungszentrum Jülich [3]. Many details of the storage ring properties relevant to these studies have been discussed previously by Benati [4].

The deuteron momentum was selected to produce a relatively clean signal of elastic scattering on carbon in the double-layered outer scintillator array of the EDDA (Elastic

Dibaryons, Dead or Alive?) detector [5,6]. This allowed the EDDA detector to operate as a polarimeter for deuteron (spin-1) vector polarization in a mode that combined neighboring scintillators into a simple, four-quadrant readout pattern, as shown in Fig. 1. The triggering scheme, as discussed below, counts single particles scattered from the target. The bars (running parallel to the beam) segment the particle readout into 32 steps in azimuthal angle. Similarly, the rings (running around the bars) provide information in steps of scattering angle. By grouping the readout signals into four quadrants (shown by colors in Fig. 1), it was possible to generate separate triggers for scattering into the left, right, down, and up directions. The asymmetry in the counting rates between opposite quadrants provided a measurement of the vertical (left versus right quadrants) and sideways (down versus up quadrants) beam polarization components as shown by Eq. (1)

$$\varepsilon_Y = \frac{3}{2} p_Y A_Y = \frac{L-R}{L+R}, \qquad \varepsilon_X = \frac{3}{2} p_X A_Y = \frac{D-U}{D+U} \qquad (1)$$

where the asymmetries, $\varepsilon_Y$ and $\varepsilon_X$, are represented in the Cartesian system by the product of the appropriate component of the beam polarization, $p_Y$ or $p_X$, and the analyzing power, $3A_Y/2$. The asymmetry (which is the same in both cases for a symmetrical detector) may be determined by comparing the counting rates in the left (*L*) and right (*R*), or the down (*D*) and (*U*), quadrants.

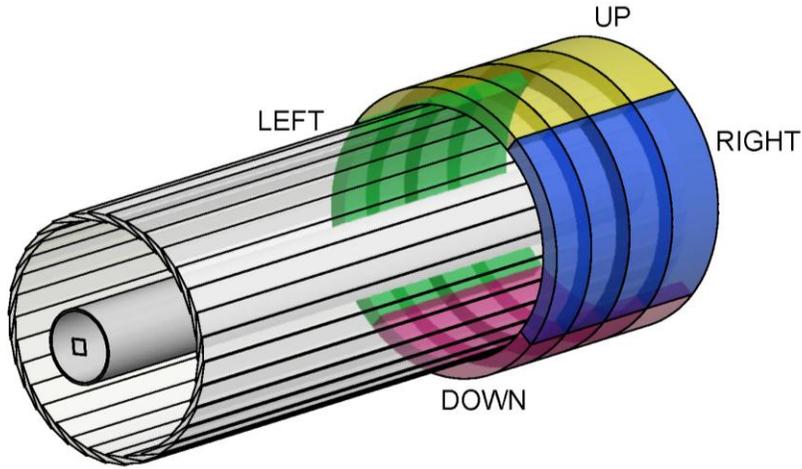

*Figure 1: Perspective drawing of the EDDA outer scintillator detectors used here as a polarimeter. The inside layer consists of 32 long bars arranged into a cylinder running parallel to the beam outside the beam pipe. The outside layer consists of a series of rings that wrap around the bars and that each intercept a narrow range of scattering angles from the target. The beam goes from left to right. Only the rings used in this experiment are shown; other surround the bars upstream. The bars are read out by photomultiplier tubes located at the ends of the bars. The rings are split into left and right halves. Each half is connected to a light guide and single photomultiplier tube. A tube target, shown here as a small open square, is located at the front of the beam pipe. Readout organization into four quadrants, denoted by the colored ring segments, is discussed in the text.*

The polarimeter target in front of the EDDA scintillators consisted of a carbon tube 17 mm long with a rectangular interior opening for the beam. In order to generate a continuous readout of the beam polarization, the beam was "extracted" slowly onto the target. This began by positioning the beam within 3 mm of the top edge, then raising the beam using steering magnets on either side of the target. The ramping rate was adjusted to spread the events reaching the EDDA scintillators throughout the time of the beam store.

Each machine storage cycle began with a vertically polarized deuteron ($\vec{D}^-$) beam that was strip-injected into the COSY ring and accelerated to 0.97 GeV/c. In the early part of the cycle, electron cooling ran for 30 s to minimize the momentum spread and transverse size of the beam. The next 10 s provided a time window in which white noise (with a frequency spectrum spanning a harmonic of the horizontal betatron tune) could be applied to a pair of horizontal electric field plates. This varying field had the effect of enlarging the horizontal size of the beam depending on the amount of power being applied. This process provided a horizontally wide beam of variable width suitable for use in the exploration of sextupole field effects on the spin tune spread. In a similar fashion, the contribution of beam size in the other dimensions to the depolarization may be individually measured and corrected. After heating, the signal for the data acquisition start was issued, ramping of the beam toward the edge of the tube target was started, and the RF-solenoid began the process of precessing the polarization into the ring plane.

*B. Time-marking system and initial data processing*

Electronic signal processing of the information from the EDDA detector scintillator array is illustrated in Fig. 2. A trigger signal was generated whenever coincident signals of a sufficient size were present in at least one bar and one ring detector of the EDDA system. Groups of 8 bars identified the four quadrants: left, up, right, and down. The four downstream ring detectors, which run perpendicular to the bars on a path around the cylinder, selected a range of polar scattering angles between $\theta_{LAB} = 9.0°$ and 14.4°. The trigger threshold for bar-ring coincidences was set to encompass the elastic scattering group based on a total summed pulse-height signal for the bar and ring scintillators. In addition, discriminators set on the individual bar and ring inputs checked that the signal was in fact a coincident event in both sets of scintillators. A triple coincidence with a high level for the summed signal was used to restrict the recorded events to be mostly deuteron-carbon elastic scattering.

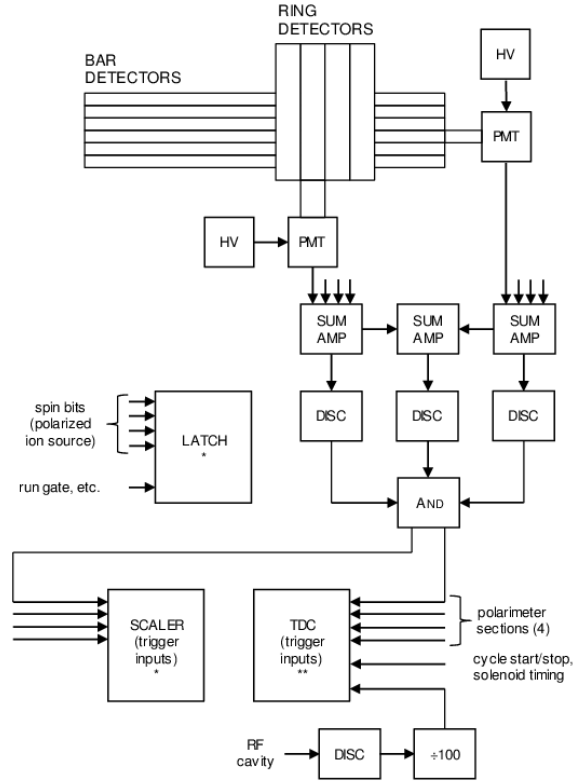

*Figure 2: Block diagram of the electronics for recording information from the EDDA detectors and other experimental parameters. Only one photomultiplier tube (PMT) is shown for the bar and ring detectors. Other PMT analog signals are summed (SUM AMP) and sent to single channel analyzers (DISC) to produce logic outputs that mark pulses above threshold. Only one of the four event streams is shown. The triggers are generated by the AND gate and recorded simultaneously in a scaler and in a time-to-digital converter (TDC). Other system inputs include a prescaled signal from the COSY RF cavity, bits to identify the polarized ion source state, and timing information associated with the machine cycle. The latch and scaler are read out periodically (400 time per second) through a CAMAC system (\*) while the TDC continuously streams data through a VME crate (\*\*).*

The trigger thresholds were adjusted to optimize the counting efficiency while maintaining a large analyzing power. In this application, only two polarization states were used, positive and negative vector polarization with no tensor polarization, along with an unpolarized state. The average polarizations were $p_Y = +0.44 \pm 0.03$ and $-0.29 \pm 0.02$ for the two vector polarized states. The analyzing power of the events accepted by the EDDA detectors is this configuration was typically $A_Y = 0.37 \pm 0.02$. The errors in the polarization and the analyzing power represent the uncertainty in the calibration of the Low Energy Polarimeter [7] located in the transfer line between the JULIC cyclotron and COSY.

Each of the four triggers was fed to a ZEA-2 GPX time-to-digital converter (TDC) constructed locally by the Forschungszentrum Jülich GmbH electronics group. The clock time period, which may be set externally, was 92.59 ps. The clock counter reached full scale at 6.4 μs, at which point an overflow bit was sent to a 20-bit register, allowing a full range up to about 6.7 s. The COSY RF-cavity signal (operating on harmonic $h = 1$ at 750602.5 Hz) was divided by 100 and used as a fifth TDC input. Besides calibration of the TDC time base against the RF-cavity oscillator for the purpose of counting beam turns, this regular input allowed any filling of the overflow register to be noted so that subsequent TDC values could be corrected to provide the right clock times. A sixth trigger was generated based on a pre-scaled signal from the RF-solenoid oscillator. To the extent that this was tuned to match the $1 - G\gamma$ RF-solenoid spin resonance, it provided an independent check for the spin tune frequency. The time associated with each trigger was passed to the analysis computer as it was generated. The RF solenoid made possible manipulation of the polarization direction, and in particular was used to rotate that polarization into the horizontal plane for further study within each machine cycle.

In parallel, an older-style data acquisition process also ran. In this case, triggers from each EDDA quadrant were counted in separate scalers. The scaler totals were read out 400 times per second. This process was used primarily to monitor online the left-right asymmetry associated with the vertical polarization. This ran throughout the beam storage time and gave polarization information prior to the start of the RF-solenoid.

*C. Online data processing*

The event file generated by the data acquisition system consisted of a list of clock times associated with each of the four polarimeter triggers, as well as time markers from the COSY RF-cavity and critical points in the machine cycle. A number of checks were made. First, clock times were adjusted so that they increased uniformly from the start time ($t_0$) despite overflows in the TDC clock value. Second, the RF-cavity times were checked to make sure that no pulses were missed and it was possible to get an accurate conversion from clock time to turn number. In most cases runs of 1-2 hours contained no missed RF-cavity pulses. In the few cases where one was missed, pulses at earlier and later times were used to make the conversion from time to turn number.

Each machine cycle begins with a vertically polarized beam. At a given moment after $t_0$, an RF solenoid is ramped on at the $1 - G\gamma$ harmonic of the spin tune, $G\gamma$, the ratio of the polarization precession frequency to the beam's orbit frequency. This harmonic, which is of the form $m \pm nG\gamma$ where $m$ and $n$ are integers, was chosen to match the optimum frequency of the RF-solenoid driver circuit. If the solenoid frequency is on the $(1 - G\gamma)f_{CYC}$ resonance within a tolerance of about 1 Hz, then it will precess the vertical polarization toward the horizontal plane, at which time the solenoid strength is ramped to zero. (Alternatively, the solenoid frequency can begin near (typically within 10 to 100 Hz) the resonant frequency $(1 - G\gamma)f_{CYC}$ (about 871434 Hz), where $f_{CYC}$ = 750602.5 Hz is the cyclotron frequency, and ramp until the center of the resonance is reached and the vertical polarization has precessed into the horizontal plane [4]. Ramping speeds are typically 5 to 40 Hz/s with a peak RF-solenoid strength of $> 2 \times 10^{-5}$ rev/turn, more than enough to completely flip the polarization.) This preparation, adjusted in advance to minimize the final vertical polarization, is initiated with a start signal that is also passed to the data acquisition as an event with a clock time. For each machine cycle, this is taken as the data acquisition start time. The RF-solenoid process is assumed to be reproducible on each cycle. Online evidence for this is the final value of the average vertical polarization component, which reproduces from cycle to cycle within the typical statistical error of 0.004.

From the start time $t_0$ onward, each transit of the beam bunch past the polarimeter target is counted as one turn in the revolution of the beam around the ring. The RF-cavity signal is pre-scaled by a factor of 100 to reduce the processing load and data file size passed to the data acquisition system in order to continuously calibrate the conversion of clock times into turn numbers associated with each polarimeter event.

The flow of the analysis from $t_0$ onward is summarized in a chart shown as Fig. 3.

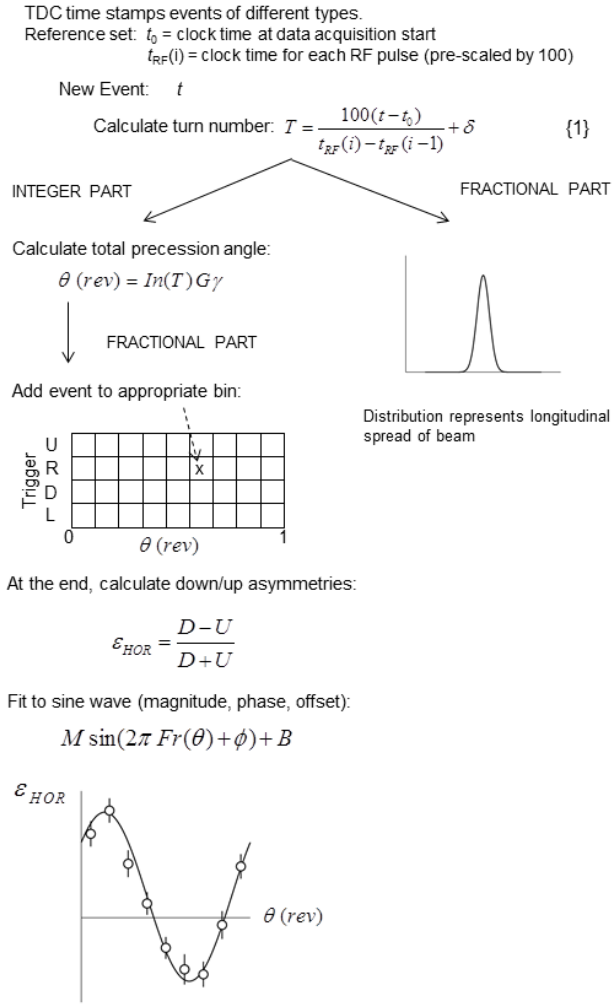

Figure 3: Flow chart of the analysis scheme. See text for details. "In" refers to the integer part of the angle $\theta$ and "Fr" refers to the fractional part. The value of $\delta$ in equation {1} is chosen

to maintain a single value for the integer part of the turn number for all parts of the beam bunch (harmonic $h = 1$). See the text for a complete description.

A starting time for the data analysis was recorded as the clock time $t_0$. Turns were counted from this point forward. Every clock time generated a real turn number $T$ (see Fig. 3). A preset phase $\delta$ was included so that the events from the passage of a single bunch by the polarimeter were all associated with the same integer part of the turn number. The separation of the turn number into the integer part and the fractional part gave access to two different sorts of information. The fraction corresponds to a measurement of the longitudinal position of a particle when it hits the EDDA target compared to the COSY ring circumference (183.4 m). This fraction was used to generate a spectrum of the beam longitudinal profile which changes as a function of the time in the store. One example is shown in Fig. 4.

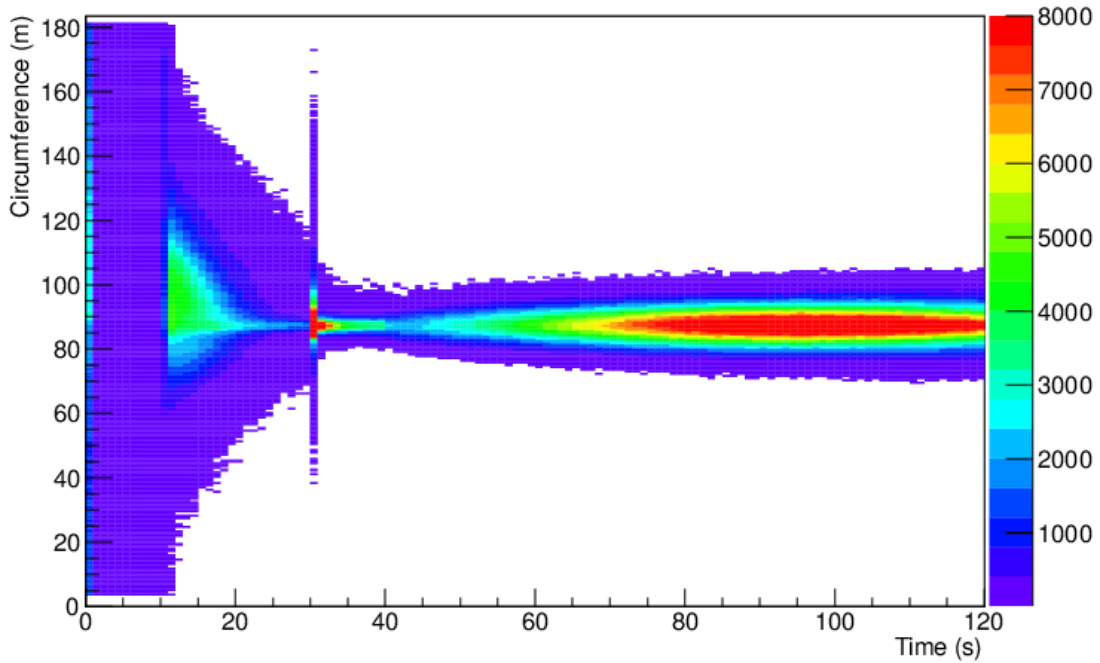

*Figure 4: The longitudinal distribution of polarimeter events for one machine cycle shown as a function of the circumference of the ring and as a function of the time during the store. The sequence of features is described in the text. A lower cutoff has been imposed on this picture at the level of 5 events/bin; below this level no events are shown. Backgrounds in the vacant regions of this figure were typically 0.1-0.2 events/bin, which is approximately $10^{-4}$ of the average rate in the colored region of the figure.*

The time in the cycle of Fig. 4 begins with the injection of the beam into COSY. After ramping the energy, the first 30 s are devoted to electron cooling. Beam bunching begins at 11 s, after which it is possible to see the effect of cooling on the longitudinal size of the bunched beam. The phase $\delta$ of the fractional part of the turn number has been adjusted to place the center of the cooled distribution near the center of this graph. Tails of the electron-cooled beam distribution send particles into the carbon polarimeter target, generating events during this 30-s period. At 30 s, the vertical beam position is moved to put the beam within 3 mm of the top inside surface of the carbon tube target. This generates a flash of particles seen in the figure as a wide, but brief, band. During the next 10 s, the beam is heated horizontally using white noise applied to horizontal electric field plates. No longitudinal growth of the beam occurs during this time. At 40 s, the RF-solenoid is energized to rotate the polarization into the horizontal plane and the final, slow extraction ramp is started. The slow ramp feeds most of the rest of the beam into the target with the core appearing near the end of the store. Since the electron cooling is off, there is some increase in beam size during this extraction time, but most of the increase in the longitudinal profile size comes from the greater sensitivity to tails of the distribution during extraction. The quality of the definition of the features of this plot would be consistent with a resolution that is less than 2% of the ring circumference. In time this would be less than 25 ns. The integrity of the central ridge in this plot confirms the choice of the phase $\delta$ in the equation {1} at the top of Fig. 3.

A first approximation to the spin tune $\nu_S$ may be obtained from the measurement of the RF-solenoid spin resonance frequency $f_{RES} = f_{CYC}(1-G\gamma)$ [2] as:

$$G\gamma = 1 - \frac{f_{RES}}{f_{CYC}} \quad (2)$$

The factor $G\gamma$ is multiplied by the integer part of the turn number to produce the total angle of the polarization rotation in the horizontal place since $t_0$. The use of the integer part recognizes that each passage of a particle past the polarimeter represents an increase of one turn number from the last passage, regardless of whether the particle happens to be ahead of or behind the main part of the beam bunch. This angle is expressed in revolutions as no factor of $2\pi$ has been included in the formula in Fig. 3. The fractional part of this angle provides the direction in which the polarization should be pointing in the horizontal plane on a scale from 0 to 1.

The next step is to pick out those angles where the polarization is pointing sideways to the beam since the down-up asymmetry generated by this polarization is a maximum (either positive or negative) there. The polarization direction at $t_0$ is unknown. So a systematic search is made by first dividing the range from 0 to 1 into 9 angle bins, sorting the events into direction bins by trigger quadrant, and then looking in each case to see which direction bins have the largest down-up asymmetry. In fact, the asymmetries for all direction bins are calculated and the nine asymmetries reproduced by a sine wave with an adjustable magnitude, phase, and offset. This is shown schematically in Fig. 3. The magnitude of the horizontal polarization is given by the ratio of the sine wave magnitude $M$ divided by the analyzing power $2A_Y/3$. If there is a systematic difference between the acceptance of the down and up detectors, it will appear as a non-zero offset in the sine wave fit as it should produce the same false asymmetry for all directions.

The fractional time width of one of the nine total angle bins is narrow, about $9 \times 10^{-7}$ after recording data for only one second. The spin tune cannot be estimated with this reliability from the values of the spin tune and RF-cavity frequencies, which are typically known only to about ± 1 Hz [4]. For the analysis shown in this paper, a central value of the spin tune was chosen to be $\nu_S = 0.160975$ based on measurements sensitive to the location of the resonance. Then a scan was made spanning ± $5 \times 10^{-6}$ in 200 steps of $5 \times 10^{-8}$ each. For each of these values of the spin tune, the asymmetry data from the one second of recording (or each machine cycle time bin as chosen in the analysis) was re-analyzed and fit to the function:

$$f(\rho) = M \sin(2\pi\rho + \varphi) + B \tag{3}$$

where $\rho = Fr(\theta)$ is the fractional part of the total precession angle. The largest value of $M$ found among these searches was chosen as the magnitude of the horizontal asymmetry for that machine cycle time bin. Examples of such scans are shown in Fig. 5 for successive 4-second time intervals during one machine cycle. The peak represents the value of the spin tune that allows all of the polarimeter events that carry information on the sideways polarization component to be sorted into the same direction bin(s) so that the size of the largest down-up asymmetry can be calculated. The position of this peak reproduces well from one machine cycle time bin to the next.

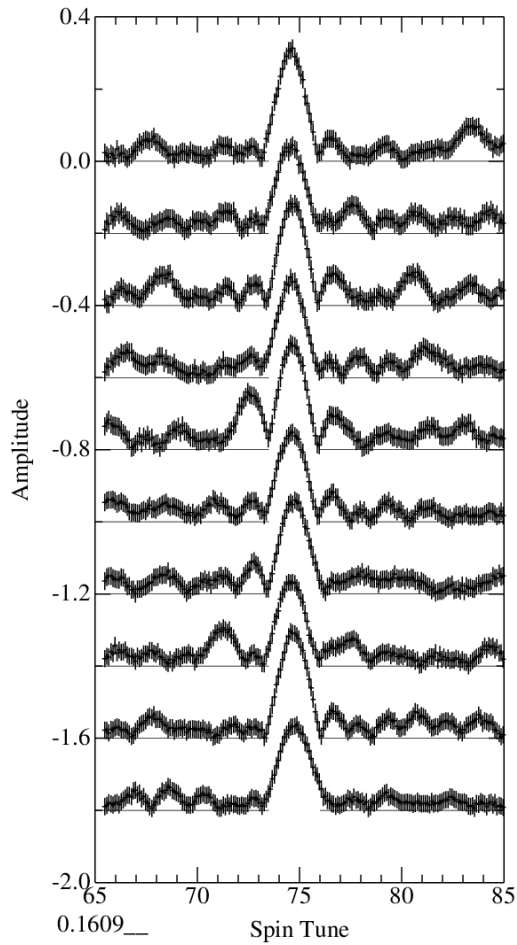

Figure 5: *Values of the magnitude of a fit of Eq. (3) to a set of events collected during successive 4-s time bins during one machine cycle time as a function of the spin tune $\nu_S$. Each successive time bin is plotted with a zero line that is shifted down by 0.2. The peak center represents the best choice of spin tune for each time bin. The typical FWHM of the peak is 1.7 ×*

$10^{-6}$ or 10 ppm. For simplicity, the first four digits of each value on the horizontal axis are noted in the lower-left corner. The value at the center of the axis is 0.160975.

In the analysis, a second, more detailed scan is made of the central 10% of this spin tune range in order to locate the best value for each machine cycle time bin. Based on the example shown in Fig. 5, a finer grid 200-step scan was made from 0.16097442 to 0.16097642. For one of the machine cycles during this run, the time sequence of best spin tune values is shown in Fig. 6. In this example, the spin tune is seen to decrease steadily during the machine cycle across values that are close to the bottom of the narrow scanning range. When this behavior is represented by a straight line (dashed line in Fig. 6), the RMS fractional deviation of these points about that line is $2 \times 10^{-8}$. Using this as an estimate of the error in the determination of the spin tune, the downward slope of the straight line is known with an error that is $1.5 \times 10^{-10}$ /s. These very small error values illustrate what is possible to achieve with this system in the study of spin effects associated with the decoherence of a polarized beam. The changes in the spin tune are smaller than the width of the spin tune peak as noted in the caption to Fig. 5. This level of precision is also smaller than the $10^{-9}$ needed in a feedback system in order to ensure that the polarization remains close to the direction of the velocity during an EDM-ring beam store lasting hundreds of seconds. While it is beyond the scope of this paper to examine possible sources of the changes in spin tune with time, second-order spin tune variations within the beam that lead to horizontal polarization loss may appear at the level of $10^{-8}$.

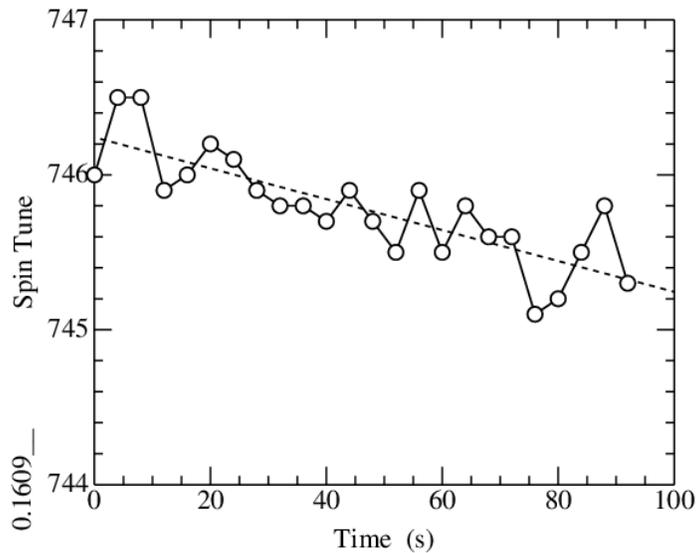

*Figure 6: Values of the spin tune chosen by the analysis program for each of a series of 4-second time bins in a machine cycle. For simplicity, the first four digits of the spin tune are noted in the lower-left corner, so the Y-axis ranges from 0.1609744 to 0.1609747. The individual points are connected with a line to guide the eye. See the text for further discussion of the dashed trend line and the scatter in these values.*

Samples of horizontal polarization data will be discussed in the next two sections.

## III. Systematic Errors in the Horizontal Polarization Magnitude

Always choosing the largest magnitude in a scan may result in horizontal polarization values that are systematically too large. But once all of the time bins in a machine cycle have been analyzed in this way, another option becomes available. It is possible to fix the value of the spin tune and re-analyze the data for each of the time bins. It then becomes clear that while the spin tune is not stationary within a storage ring beam cycle, the direction of the maximal polarization changes in a nearly smooth fashion that is illustrated by changes in the optimum phase $\varphi$ as shown in Fig. 7. Here the figure depicts nine successive machine cycles. By locking the phase of the RF-solenoid and the RF-cavity together at the start of each machine cycle, we have managed to get each phase history to begin near 1 rad in all cases. (In the figure, these starting points were shifted negatively by 2 rad for each successive case to separate the curves.) After that the phase follows the particular history of that beam storage cycle. A rising phase indicates that the chosen value of the spin tune is slightly too small. The correction varies as the slope of the phase history. But as the storage time proceeds, the curve bends and heads down again. Thus the actual spin tune starts above the chosen value and, during the store, moves downward, ending below the value (as shown in Fig. 6). In many cases, the final phase ends up close to its starting value. From the beginning to the end, the largest slope of the typical phase curve changes by about 0.13 rad/s. This leads to a total change in the spin tune from the beginning to the end of about $2 \times 10^{-7}$, which is about twice the change shown in Fig. 6 directly from the spin tune itself, a change that may simply reflect the fact that the two sets of data were taken at different times. While individual machine cycles may vary, the size of this change remains similar across many machine cycles and runs if the machine operating conditions are unchanged. The error bars reflect the statistics available in this run.

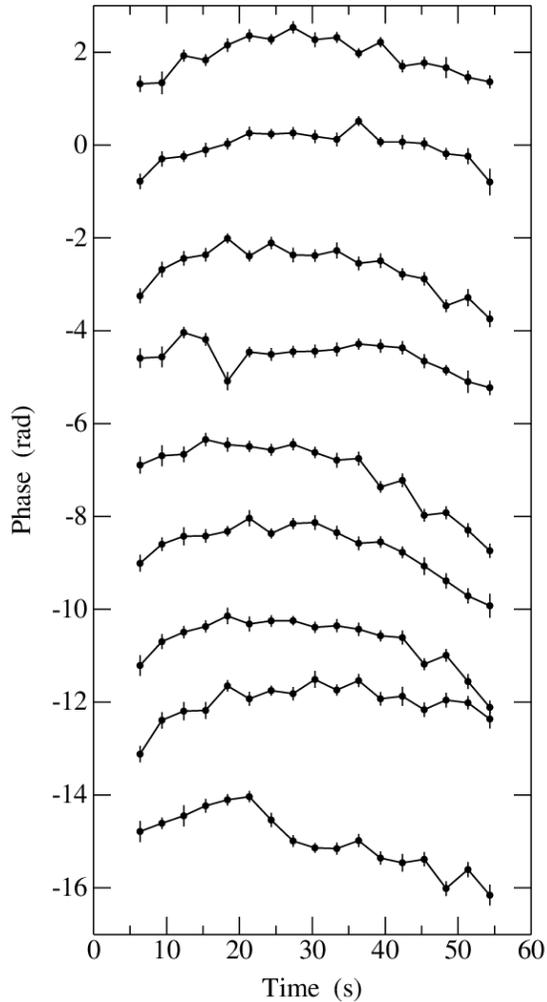

*Figure 7: Measurements of the phase $\varphi$ in Eq. (3) as a function of the time during a machine cycle in seconds. The size of the machine cycle time bins is 3 s. The spin tune was fixed at a single value for this analysis. The vertical units are radians. The nine different curves show the behavior for subsequent machine cycles during a typical run. In each successive case, 2 rad has been subtracted from the phase values to separate the curves. Each curve begins near 1 rad and then follows its own history.*

An examination of the time dependence of the phase also reveals whether or not it makes sense to assume that there is little smearing of the horizontal polarization signal in a single time bin, as would be the case if the phase changes significantly from one bin to the next bin. This can be checked by choosing slightly different spin tune values and observing the impact of this choice on the history of the phase. In this sense, the best choice would be the one that produces the least phase variation with time. That choice is "correct" only for that part of the machine cycle where the time derivative of the phase is zero. A refinement of this analysis procedure might involve choosing a smooth function rather than a fixed value to represent the spin tune, thus rendering the phase roughly constant with time. Such a representation would need to be made for each machine cycle of every run and is not addressed in this discussion.

Figure 8 shows three results from an analysis of a set of horizontal polarization data. The analysis with a fixed spin tune, whose phase variation is shown in Fig. 7, is illustrated in Fig. 8 with the black, solid points. If the highest magnitude is chosen in each case regardless of the spin tune or phase value, then the magenta points are the result. These points are significantly higher than the black points (about two standard deviations). If the spin tune is changed by $2.7 \times 10^{-7}$ so that the phase varies by about 22 rad during the machine cycle, then the spin tune is smeared within each time bin and the lower asymmetry shown by the blue points is the result. Thus the analysis is sensitive to the choice of spin tune and the flexibility allowed in the fits to each set of asymmetry measurements. Fixing the spin tune at a value that minimizes the phase variation appears to be the best choice because it avoids the effects of both statistical fluctuations in choosing the largest magnitude and the phase smearing from choosing the wrong spin tune.

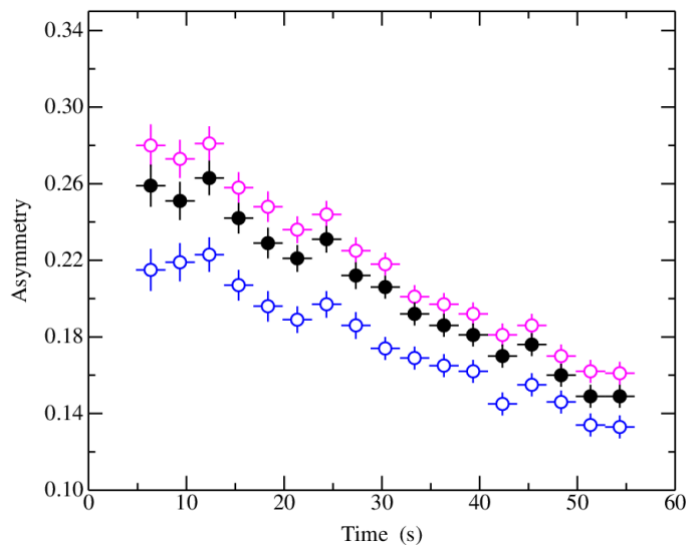

*Figure 8: Measurements of the magnitude of the horizontal polarization asymmetry as a function of the time during the data-taking portion of the machine cycle. The black points were analyzed using a fixed spin tune. The magenta points show the largest value of the magnitude when the spin tune value is allowed to vary from bin to bin. If a poor choice is made for the spin tune in the fixed tune analysis and the phase changes by 20-30 radians during the machine cycle, then there is smearing in the time bins which results in a smaller fitted magnitude, as shown by the blue points. Note the suppressed zero on the vertical axis.*

Raw data from different machine cycles may be combined only if two conditions are met: (1) the relative phase between the RF-cavity high voltage and the RF-solenoid is fixed at the starting point, and (2) the magnetic field of the ring following the synchrotron ramp is highly reproducible. Condition (1) was met by creating a discriminator signal from each of these sinusoidal signal generators and waiting to start the RF-solenoid process until these signals were in coincidence within a time window of 40 ns. This produced a situation where the starting phase at the beginning of graphs such as those shown in Fig. 7 was always the same (about 1 rad). But the subsequent history of the phase on different machine cycles was different, so condition (2) was not met to a degree that would allow the raw event data to be added between machine cycles. This meant that the analysis just described had to be completed for each machine cycle independently. At the end, the magnitudes for each time bin may be averaged, weighted according to the statistics available for each cycle. This places a premium on having an adequate event rate from the EDDA detectors and choosing the time bin width so that within each time bin there are a sufficient number of events to make the analysis statistically meaningful. For Fig. 8, an error of 0.02 in the magnitude would require a few thousand events in each bin.

*C. Positive bias in the measurement of small horizontal polarization magnitudes*

The positive bias that results from always choosing the largest amplitude point from a spin tune scan has been discussed. This may be addressed by freezing the choice of the spin tune in the fits.

In Fig. 5 it can be seen that away from the spin tune peak the values of the magnitude appear to scatter about a positive average rather than zero. This reflects a general property of using Eq. (3) to represent the asymmetries measured for different directions in the horizontal plane. If the magnitude and the phase are both allowed to vary in the fit, a positive magnitude will always be the result, even if there is no signal and the asymmetries are randomly distributed about zero. The size of this positive bias depends on the amount of scatter in the data, which itself reflects the statistical error in the asymmetry in each direction.

A Monte Carlo calculation of this effect is shown in Figs. 9 and 10. The projection plot shows along the observed asymmetry axis the magnitudes generated for some assumed real asymmetry. In each case a set of nine direction bins was considered. A sine wave with a variable amount of real asymmetry was used and the data points were distributed randomly in accordance with a standard deviation of $\sigma = 0.25$ about the sine wave. At the highest values of the real asymmetry (~0.5), the observed values are distributed near 0.5 as well with a nearly Gaussian shape. However, for values of the real asymmetry near zero, the magnitudes are shifted systematically upward. The chance of obtaining a zero magnitude vanishes. At zero real asymmetry the mean value of the distribution is given by $\sigma\sqrt{\pi/N_{BINS}} = 0.14$ (see the Appendix for a derivation). The most probable value is about 0.11. Once the real asymmetry becomes greater than the average value at zero asymmetry, the average magnitude tends quickly toward the real value.

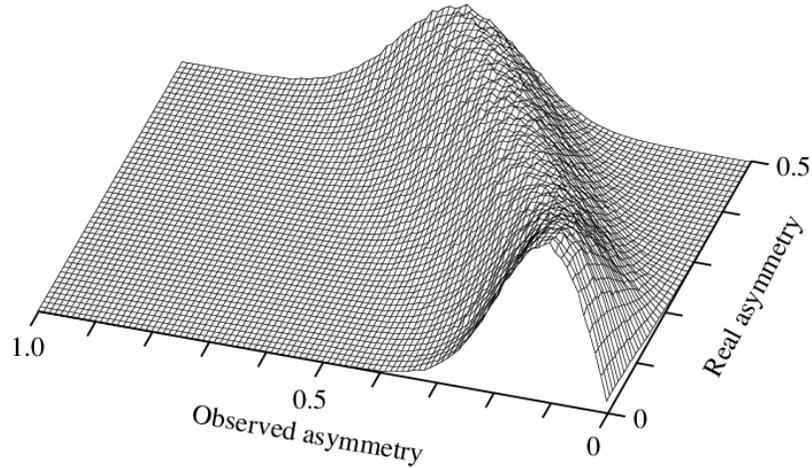

*Figure 9. This projection shows the distribution of magnitudes from a Monte Carlo calculation based on Eq. (3) for a set of nine direction bins, each with a statistical error of 0.25 and varying amounts of real asymmetry.*

These trends are also represented in Fig. 10 where the mean observed asymmetry is plotted as a series of dots as a function of the real asymmetry. A perfect response with no positive bias is shown by the straight diagonal line. The curved line is a hyperbola with the mean observed asymmetry with no signal as the minimum point and the straight line as the asymptote; this illustrates that such a simple function is not an accurate quantitative representation of the Monte Carlo result.

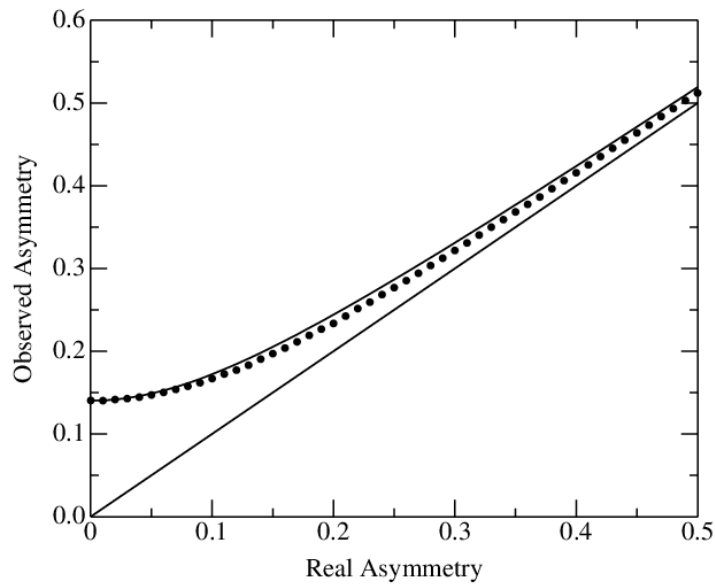

*Figure 10: The dots represent a Monte Carlo estimate of the average observed asymmetry for an analysis based on Eq. (3) with a free magnitude and phase. In addition to the random scatter due to a statistical error on each asymmetry of 0.25, there is a real asymmetry added according to the X-axis value. In the absence of a positive bias, the dots should lie along the straight line*

*where observed and real asymmetries are equal. The curved line represents a hyperbola with the correct endpoint at zero real asymmetry, but it fails to match the Monte Carlo simulation.*

Because it is possible to have average magnitudes in individual machine cycles that are less than the average value for a zero real asymmetry, it is not practical to invert this function to create a "corrected" set of magnitude measurements. As developed in the next section, we have chosen to start with template shapes that are constructed to represent the time dependence of the horizontal polarization and apply the positive bias to them, constructing another function to match the measurements.

The positive bias issue arises when both the magnitude and the phase of a fit (as in Eq. (3)) are allowed to be freely adjustable, especially when there are a relatively modest number (~10) of bins in the horizontal plane direction. This is always the situation when a new set of measurements is examined for the first time. If a way existed to independently determine the phase, based either on additional information about the running conditions or some smooth trend through the phase measurements, then this effect would no longer exist. The phase measurements shown in Fig. 7 illustrate that any empirically-determined smooth trend will be unique to each machine cycle. More importantly, when the horizontal polarization becomes small, the positive bias correction becomes more significant. Information from the data about the trends illustrated in Fig. 7 tends to disappear as the phases for individual time bins start to vary randomly between 0 and $2\pi$. As additional data of this sort are taken, more information will become available concerning the systematics of phase variation and its connection to trends in the values of the spin tune. For now, we will employ this model of the positive bias as a correction to a template for the horizontal polarization time dependence. This is discussed in the following section.

### IV. Extraction of the spin coherence time from horizontal asymmetries

*A. Introduction*

For the purpose of exploring the effect of higher-order (sextupole) magnetic fields on the longevity of polarization in a storage ring such as COSY, it is necessary to have a way to define the spin coherence time or horizontal polarization lifetime and to provide a procedure through which it may be extracted from horizontal polarization measurements. This problem is made more difficult because the time dependence of the horizontal polarization does not follow a single shape, but depends on the mechanism by which the polarization is being lost. This makes any choice of a definition of the spin coherence time arbitrary. Our resolution to these issues is to create a library of template shapes for which the polarization declines following different patterns. The one used for any given set of measurements is the one that provides the best reproduction of the time-dependence of the horizontal polarization measurements. The value for the polarization lifetime will be defined as the time at which the polarization has fallen by a factor of exp(–0.5), a definition that matches the width of a Gaussian function. This choice facilitates comparison to other beam parameter estimates based on Gaussian distributions.

In this paper, we introduce a set of templates based on a model of polarization loss due to finite emittance of the beam. The shapes are arranged into a continuum that permits selection of the correct choice based on an algorithm. The shapes are not produced from a functional form, but are generated using a computer modeling procedure. So it is necessary to produce a digital library of shapes from which the best is selected. Each shape is matched to a set of magnitude measurements by scaling the vertical and horizontal dimensions of the template until a best representation is achieved using a least squares criterion.

For the initial study using the time-marking system, the beam at COSY was prepared in a special way to emphasize only one contribution to the depolarization, the horizontal emittance. The first step in beam preparation was to electron-cool the beam, shrinking both horizontal and vertical emittance as well as the momentum spread associated with the synchrotron oscillations of a bunched beam. About 30 s were needed to achieve equilibrium in the cooled beam [4]. Then the cooling was turned off and white noise was applied to a pair of horizontal electric field plates for 10 s to increase only the horizontal emittance. This allowed studies to be made of how the decoherence from only this source could be reduced.

In the remainder of this section, the procedure for creating templates suitable to this situation is described, and then the analysis process is demonstrated using as examples a set of runs with different polarization lifetimes.

*B. Definition of the templates*

For a bunched beam, depolarization may be driven by the lengthening of the particle path associated with oscillations about the reference orbit. Vertical and horizontal oscillations are usually treated independently and typically have different tunes or oscillation frequencies. In either case, the size of an oscillation may be characterized either by the maximum distance from the reference orbit that may be reached at a particular point along the ring, or by the maximum angle of the particle track as it crosses the reference orbit. In this development, we will make use of the track angle.

The population of oscillating orbits is typically treated using a Gaussian distribution whose widths are represented by either $\sigma_X$ or $\sigma_Y$, with $X$ and $Y$ being the coordinates in and perpendicular to the ring plane. To make a simple simulation of this situation, the starting point is a set of values of the maximum angles, $\theta_{X,i}$ and $\theta_{Y,i}$, that are chosen from the two Gaussian distributions with pre-selected widths, $\sigma_X$ and $\sigma_Y$. The index $i$ merely counts the members of these sets. In general, the elements of the two sets, $\theta_{X,i}$ and $\theta_{Y,i}$, will have different values for the same $i$. From these angles we wish to create a template characterized by the ratio of the two Gaussian widths as

$$\alpha = \frac{\sigma_Y}{\sigma_X} < 1 \tag{4}$$

No generality is lost if we assume $\sigma_Y < \sigma_X$ since the effects of the two distributions will be added and it does not matter whether $X$ or $Y$ is the smallest.

The increase in the particle path length is proportional to the set of values, $\theta_{X,i}^2 + \theta_{Y,i}^2$. As time passes, the path length change since $t_0$ grows linearly with time. For a bunched beam, the change in path length is also proportional (due to the change in the relativistic factor $\gamma$) to the change in the total rotation angle of the particle spin axis since $t_0$ relative to the value for the reference orbit. If we take *n* as a dimensionless, evenly-spaced index for the growth of time, the change in the total spin rotation angle becomes

$$\phi_{i,n} = n\left(\theta_{X,i}^2 + \theta_{Y,i}^2\right) \tag{5}$$

In a more complete model of this process, *n* would contain kinematic factors such as $\beta$ and $\gamma$ as well as factors that relate the maximum orbit angles, $\theta_{X,i}$ and $\theta_{Y,i}$, measured in one place to the path length change for the whole ring. These factors are taken into account here when the time scale, as described below, is adjusted to match horizontal polarization data. The magnitude of the polarization begins to fall significantly when *n* becomes large enough that the width of the distribution of $\phi_{i,t}$ is about 1 rad. In order to map both the regions below and above 1 rad with an adequate density of numerical points, $\sigma_X$ was taken to be 1 rad and *n* was allowed to vary from 0.01 to 20, defining each template by 2000 points. $\sigma_Y$ was chosen to be less than or equal to $\sigma_X$, thus specifying $\alpha$.

From the collection of points $\phi_{i,t}$ we must now calculate a horizontal plane polarization for each point *n*. Taking these points as angles, the polarizations may be given by

$$p_Z(n) = \frac{1}{N} \sum_i \cos\phi_{i,n} \tag{6}$$

$$p_X(n) = \frac{1}{N} \sum_i \sin\phi_{i,n} \tag{7}$$

$$p_\alpha(n) = \sqrt{p_Z^2(n) + p_X^2(n)} \tag{8}$$

where *N* is the number of Gaussian points, *Z* represents the axis parallel to the particle velocity, and $p_\alpha(n)$ is the final template for a given $\alpha$.

It is assumed that, after the RF-solenoid has precessed the polarization into the horizontal plane, all of the particle spins remain aligned with each other at the maximum polarization. Then, over time, they spread around a unit circle in the horizontal plane. The contribution of each particle spin to the total polarization is found by taking the average projections along the *X* and *Z* axes averaged over the whole particle population. Then the ring plane polarization is the sum of these two averages in quadrature, as given by Eqs. 6-8. This sum may be plotted as a function of time. An example of a distribution of 300 points after various times have elapsed is shown for the three unit circles in Fig. 11. With more time, the distribution of spins will continue to spread out, becoming more uniform, and the polarization will fall.

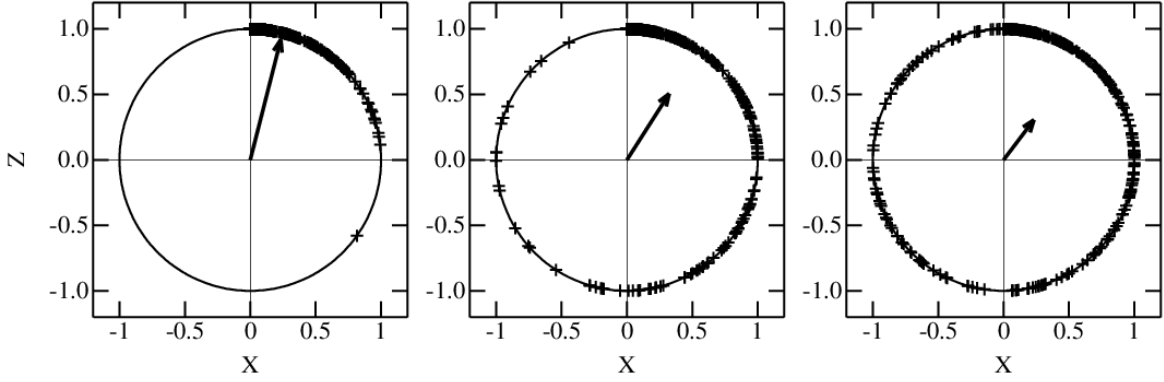

*Figure 11: A sample distribution of the points along a unit circle in the horizontal plane (X sideways and Z along the beam direction) that indicates the spin directions of an ensemble of particles for three values of the time index, n = 0.25, 1, and 4 going from the left panel to the right. The starting point at n = 0 is at the top of the graph. Since the changes in spin tune go as the square of the angles, all deviations from the reference orbit are positive and the distribution spreads in only one direction around the unit circle, shown here as clockwise. The tail of the distribution is evident as the population of spin directions becomes thinner along the circumference of the unit circle. Only one Gaussian distributions of points was used, which is equivalent to $\alpha = 0$. The arrows show the direction and magnitude of the resulting polarization. The three values of the magnitude are 0.95, 0.60, and 0.38. These values form a part of the template for $\alpha = 0$.*

The horizontal and vertical axes of the ring may have different emittance distributions. In our experimental case, the ratio of vertical to horizontal emittance sizes was deliberately varied. Such a situation leads to different polarization histories. Some sample template curves are shown in Fig. 12 as a function of $\alpha$.

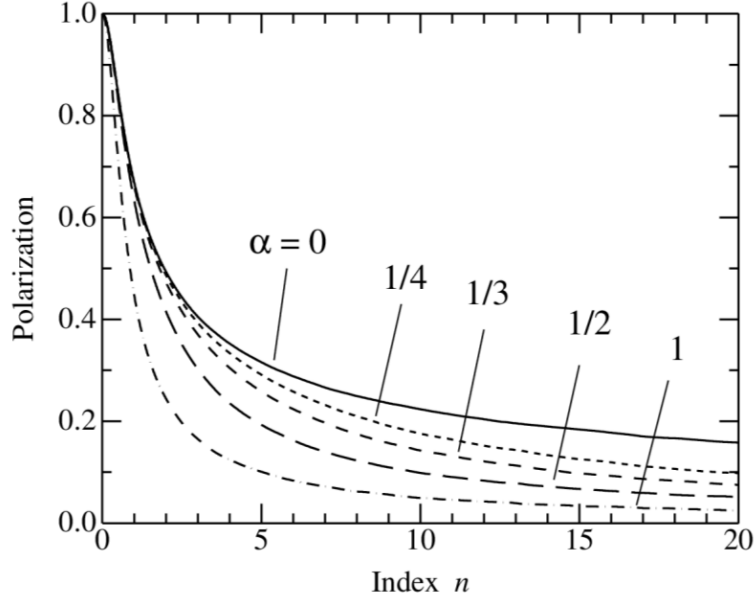

*Figure 12: Five sample templates for the time dependence of the horizontal polarization in arbitrary index units. The values for $\alpha$ are marked for each template. Values of $\alpha$ near zero represent a wide but vertically short beam. A value of $\alpha = 1$ represents a round beam. These curves include a factor of $\sqrt{1+\alpha^2}$ in the time coordinate as discussed in the text.*

The curves in Fig. 12 represent polarization shapes as a function, $p_\alpha$, of the index $n$. To scale this to data, both the polarization and the time axis must be multiplied by an adjustable factor. The start time, $n = 0$, is placed at the time when the RF solenoid has completed the rotation of the polarization into the horizontal place. This creates a new function, $f(t_{EXP})$ of the time in the experiment, that is determined by two parameters, $a_1$ and $a_2$.

$$f(t_{EXP}) = a_1 p_\alpha(n) \tag{9}$$

$$t_{EXP} = a_2 \sqrt{1+\alpha^2}\, n \tag{10}$$

The factor inside the square root is helpful in making the curves similar in the region where the polarization is above 0.8. Non-linear regression is used to choose the best values for $a_1$ and $a_2$.

At each iteration of the fitting process, the positive bias is included based on the known errors for the individual asymmetry points. An example of such a fit with the change due to the bias is shown in Fig. 13.

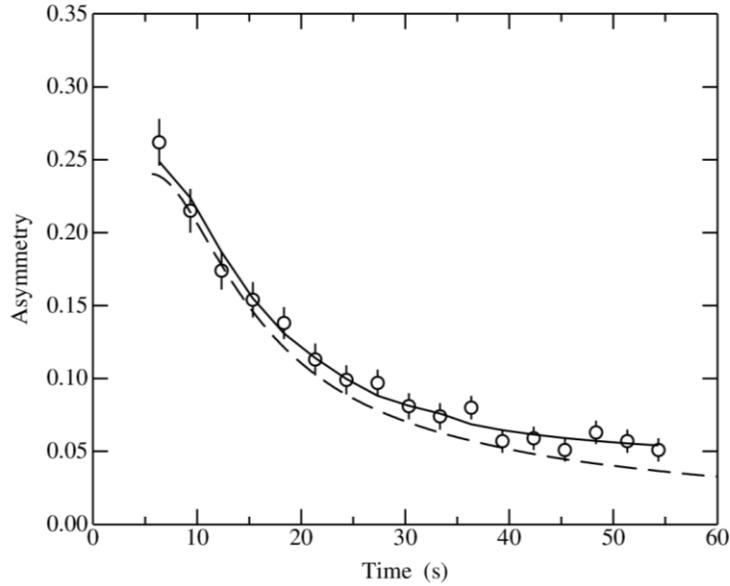

*Figure 13: Measurements of the time dependence of the horizontal magnitude (see Eq. (3)) compared with the best fit template curve (dashed line) and the template curve with the positive bias effects included (solid line). The modified curve is only defined where there are data points whose error determines the size of the positive bias.*

In practice, a series of templates with varying values of $\alpha$ are tried. The templates with the smallest value of the reduced chi square are identified. This allows an interpolation to the value of $\alpha$ with the minimum reduced chi square. From this information, it is possible to interpolate to the best value of $a_2$, the coefficient that is proportional to the spin coherence time. Such a minimization for two different spin coherence times is illustrated by the left and right series of panels in Fig. 14.

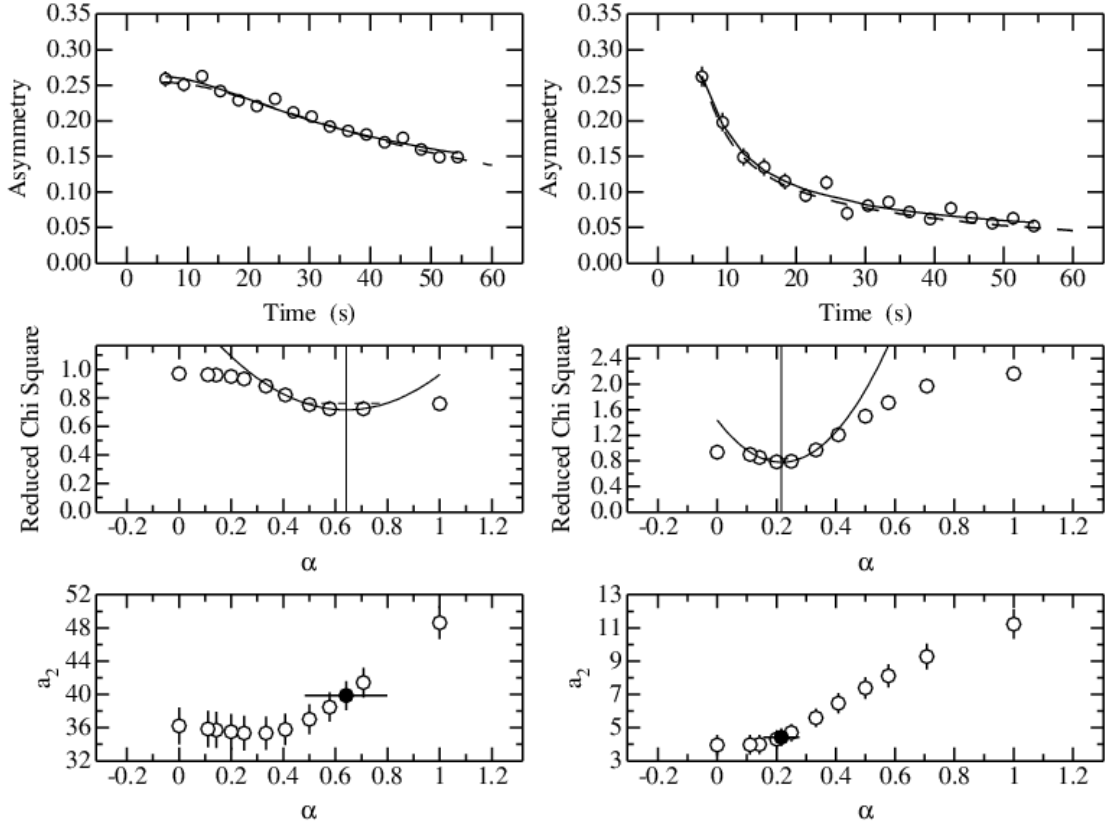

*Figure 14: Two different analysis cases are shown for a long polarization lifetime on the left and a short polarization lifetime on the right. The top panel shows the data and the fits with the same legend for the curves as described in Fig. 13. For the purpose of illustration, the template and bias-corrected curves (as in Fig. 13) with the value of $\alpha$ closest to the best fit value are shown. In the second panel is a series of reduced chi square values for each of the templates spanning a series of $\alpha$ values between 0 and 1. In both cases there is a minimum that indicates the best choice of $\alpha$. This is located more precisely by reproducing the local shape of the reduced chi square distribution with a parabola as shown. The vertical line indicates the value for $\alpha$ where the parabola has a minimum. In the third panel, each fit with a specific template also generates a value with a statistical error for the $a_2$ parameter. The value with the filled point is an interpolation of the values of $a_2$ to the best value of $\alpha$. The parabolic fit to the reduced chi square results may also be used to determine the error in the value of $\alpha$. This is projected onto the $a_2$ result as the horizontal error on the black point, and indicates the uncertainty beyond statistics from the quality of the template fit. In the left-side case, this uncertainty exceeds the statistical uncertainty. Since the beam distribution is obtained by horizontal heating, the right-side panels represent a beam with a large horizontal extent, thus leading to a small value of $\alpha$ and a short polarization lifetime. Similarly, the left-side panels are for a more nearly round beam. Thus the values of $\alpha$ correspond to the known changes in the setup of the beam. For the long and short lifetime cases (left and right panels), the values of $a_2$ = 39.8 ± 4.1 and 4.41 ± 0.72, leading to Gaussian-equivalent polarization lifetimes of 44.6 ± 4.5 s and 5.62 ± 0.92 s respectively (see next section). The errors include statistics and quality of fit.*

*C. Definition of the polarization lifetime*

      To have a standard for the comparison of polarization lifetimes, the analysis reported here adopts a particular definition based on the width of a Gaussian distribution. This choice often coordinates well with other calculations and estimates which use a Gaussian distribution to describe some other related property or process. Since none of the templates described above are Gaussian, we have chosen the spin coherence time to be the width of the distribution when the polarization has fallen to exp(–0.5) = 0.606… from a starting point of unity. This is the level at which the Gaussian distribution half width is equal to the usual width parameter $\sigma$. The connection between this definition and the shape of each template is unique to that template.

      To see this connection, an example is described for the creation of a template for a single Gaussian distribution of angles, $\theta_i$. The width of the Gaussian was chosen to be $\sigma = 1$ rad with the expectation that the spin coherence time for this distribution would be close to one index unit ($n = 1$). Next, the distribution of spin tune changes was created as the squares of the angles. As time proceeds, the elements of this distribution expand linearly around the horizontal plane. The polarization associated with any such distribution is calculated by averaging its *X* and *Y* components and then adding them in quadrature. Polarizations for other cases are calculated at earlier or later times by scaling all of the squared angle points by *n* and recalculating the polarization. As *n* starts at 0 and increases, the polarization starts at one and falls smoothly. As expected, the value of the polarization reached the spin coherence time definition value of exp(–0.5) when *n* was close to one. In fact, $n = 1.254$ (see template for $\alpha = 0$ in Fig. 12). As *n* increases further, an entire polarization decay curve is created. One way to use this decay curve as a template is to expand or contract the time scale until (with the magnitude separately adjusted) the template matches a given set of data (with the positive bias correction included). The match maps the point at $n = 1.254$ into the point representing the polarization lifetime of the decay curve data. The actual polarization lifetime value is then obtained by taking the amount of scaling (the $a_2$ parameter in Eq. (5)) and multiplying by 1.254 to correct for the amount that the template polarization lifetime differs from one. This produces a polarization lifetime in whatever units (usually seconds) were used for the data recording.

      The case for values of $\alpha > 0$ is similar, except that the original spin tune distribution is the sum of the two distributions from the appropriate vertical and horizontal emittance angle distributions. Unfortunately, the template polarization lifetime for these cases is not $n = 1.254$ but is something nearby. Part of the difference can be made up by including an additional factor of $\sqrt{1+\alpha^2}$, as it appears in Eq. (10). The remaining correction, now closer to one for all values of $\alpha$, is shown in Fig. 15. Thus the dark points in Fig. 14 are close to, but not equal to, the spin coherence times from their respective data sets.

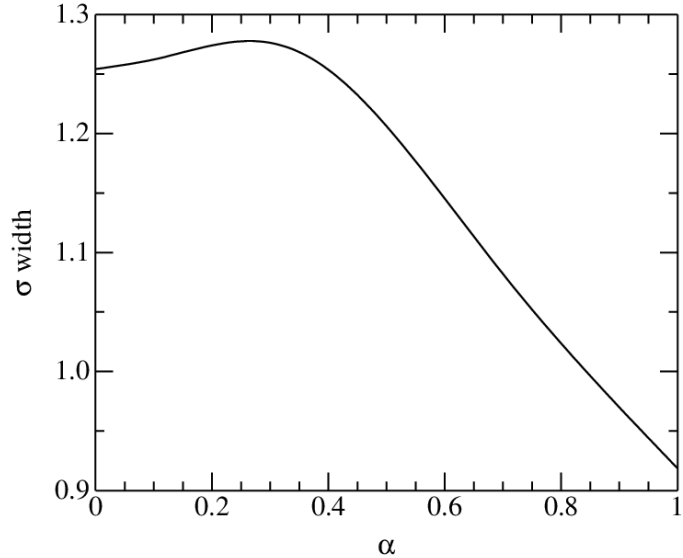

*Figure 15. Graph of the factor that converts values of $a_2$ to the time base of the measurements as a function of $\alpha$.*

### V. Conclusions

    In this paper we have described the time-marking of polarimeter events at the COSY ring with a clock time in order to unfold the rotation of the horizontal polarization and measure its magnitude. The TDC was built by the electronics group (ZEA-2) at the Forschungszentrum Jülich. With this information and with a calibration made in real time against the oscillating RF-cavity voltage, it is possible to observe the longitudinal spatial distribution of the beam and how it evolves during any machine store from the fractional part of the turn number. With the integral part of the turn number, it is possible to associate each event with a total rotation angle for the polarization, assuming a value of the spin tune. While a starting estimate of the spin tune is available from the value of an RF-solenoid induced resonant frequency relative to the RF-cavity frequency, it proved necessary to search for the best possible choice. With this value, it proved possible to track the horizontal polarization in discrete time bins during the beam store. Using the phase as measured through the machine cycle, a criterion of minimal phase change was developed to specify the spin tune used for the final extraction of the horizontal polarization.

    The data was matched to template shapes that represented the polarization loss due to combinations of horizontal and vertical emittance. In the process of developing this system, we observed a systematic error present at low asymmetries that tends to generate a positive bias in the magnitude of the horizontal polarization. A correction must be added to whatever template shape is used to describe the data. The combination could be adjusted to best reproduce the time dependence of the horizontal polarization. The spin coherence time, defined in analogy to the width of a Gaussian function, was extracted from the value of the fitting parameter $a_2$. The template shape varied depending on the ratio of the vertical to the horizontal beam emittance. The coefficient converting $a_2$ to spin coherence time also depends on this ratio.

**Acknowledgements**

The authors wish to thank other members of the Storage Ring EDM Collaboration [8] and the JEDI Collaboration [9] for their help with this experiment. We also wish to acknowledge the staff of COSY for providing good working conditions and for their support of the technical aspects of this experiment. This work has been financially supported by the Forschungszentrum Jülich GmbH, Jülich, Germany via COSY-FFE, the EU Integrated Infrastructure Initiative (FP7-10 INFRASTRUCTURES-2012-1, Grant Agreement No. 227431) and the Shota Rustaveli National Science Foundation of the Republic of Georgia. This manuscript has been authorized by the Brookhaven Science Associates, LLC under Contract No. DE-AC02-98CH10886 with the US Department of Energy.

**Appendix**

In the set of directions in the horizontal (ring) plane, let $\sigma$ be the standard deviation in the asymmetry for each of the $N$ bins. When these asymmetries are represented by Eq. (3), the result will be a sinusoidal function whose mean value is $b$. With reference to this mean value, the circle may be divided into two regions, one where $f(\rho) - b$ is positive and another where it is negative. The mean value of the asymmetries in these two regions will be $b + \delta$ and $b - \delta$ where $\delta$ is the expectation (average) value of one-half of a Gaussian (normal) distribution whose width is $\sigma\sqrt{N/2}$, the result of averaging over $N/2$ bins. If the Gaussian distribution is denoted by $F[0,\sigma]$ with a mean value of zero, then

$$\delta = \frac{\int_0^\infty x F\left[0, \sigma/\sqrt{N/2}\right] dx}{\int_0^\infty F\left[0, \sigma/\sqrt{N/2}\right] dx} = \sqrt{\frac{2}{\pi}} \frac{\sigma}{\sqrt{N/2}} \tag{A.1}$$

At the same time, one region is represented by a function, $f(\rho)$ from Eq. (3), which has an average value of

$$\langle f(\rho) \rangle = \frac{1}{\pi} \int_0^\pi a \sin x \, dx = \frac{2a}{\pi} \tag{A.2}$$

The fit equates these two quantities, yielding

$$a = \sigma \sqrt{\frac{\pi}{N}} \tag{A.3}$$

\* <stephene@indiana.edu>